\documentstyle[12pt]{article}
\def \ms {{\overline{\mbox{MS}}}}

\newcommand{\titul}[1] {\begin{center}{\large\bf #1 } \end{center}\vskip 1.cm}
\newcommand{\autor}[1] {\begin {center} {\large \lineskip .5em #1 }
                        \end   {center} }
\newcommand{\lugar}[1] {\begin{center} {\it #1} \end{center}}
\newcommand{\abstr}[1] {{\begin{center} \vskip .5cm {\bf Abstract
                        \vspace{0pt}} \end{center}}\begin{quote} #1
                        \end{quote}}

\newcommand{\ep}{\varepsilon}
\begin{document}
\begin{titlepage}
.
\begin{flushright} {\bf US-FT/3-02 \\ 
July, 2002} \end{flushright}

\vskip 3.cm
\titul{ 
Small $x$ behavior of the slope 
$d \ln F_2/d\ln (1/x)$ \\ in 
the framework of perturbative QCD
}

\autor{A.V. Kotikov
}
\lugar{Bogoliubov Laboratory of Theoretical Physics\\
Joint Institute for Nuclear Research\\
141980 Dubna, Russia}
\autor{G. Parente
}
\lugar{Departamento de F\'\i sica de Part\'\i culas\\
Universidade de Santiago de Compostela\\
15706 Santiago de Compostela, Spain}
\abstr{
Using an analytical parameterization for the behavior of the $x$ {\rm slope}
of the structure function $F_2$ at small $x$ in perturbative
QCD, at the
leading twist approximation of the Wilson operator product expansion, and
applying a flat initial condition in the DGLAP evolution equations,
we found very good agreement
with new precise deep inelastic scattering  experimental data from HERA.

PACS number(s): 13.60.Hb, 12.38.Bx, 13.15.Dk

}
\end{titlepage}
\newpage

\pagestyle{plain}

\section{Introduction} \indent

The measurement of the deep-inelastic scattering (DIS)
structure function $F_2$ \cite{H197,H101,ZEUS01},
and the derivatives $dF_2/d\ln(Q^2)$ \cite{H197,H101,Surrow} and 
$d\ln F_2/d\ln(1/x)$ \cite{H1slo,Surrow}
in HERA 
have permitted the access to
a very interesting kinematical range for testing the theoretical
ideas on the behavior of quarks and gluons \cite{Ynbook}
carrying a very low fraction
of momentum of the proton, the so-called small $x$ region.
In this limit one expects that
nonperturbative effects may give essential contributions. However, the
reasonable agreement between HERA data and the 
next-to-leading (NLO) approximation of
perturbative
QCD has been observed for $Q^2 \geq 2$ GeV$^2$ (see review Ref. \cite{CoDeRo}
and references therein) and, thus,
perturbative QCD could describe the
evolution of $F_2$ and its derivatives
up to very low $Q^2$ values,
traditionally explained by soft processes.
It is of fundamental importance to find out the kinematical region where
the well-established perturbative QCD formalism
can be safely applied at small $x$.

The standard program to study the $x$ behavior of
quarks and gluons
is carried out by comparison of data
with the numerical solution of the
Dokshitzer-Gribov-Lipatov-Altarelli-Parisi (DGLAP)\footnote{
At small $x$ there is a different approach
based on the Balitsky-Fadin-Kuraev-Lipatov (BFKL) equation \cite{BFKL}, whose
application is out of the scope of this work. However,
sometimes we use below the BFKL-based predictions in the discussions and
for the comparison with our results in generalized DAS.}
equations \cite{DGLAP}
by fitting the parameters of the
$x$ profile of partons at some initial $Q_0^2$ and
the QCD energy scale $\Lambda$ \cite{fits,GRV}.
However, for analyzing exclusively the
small $x$ region, there is the alternative of doing a simpler analysis
by using some of the existing analytical solutions of DGLAP 
in the small $x$
limit \cite{BF1}-\cite{HT}.
This was done so in Ref. \cite{BF1}
where it was pointed out that the HERA small $x$ data can be
interpreted in 
terms of the so-called doubled asymptotic scaling (DAS) phenomenon
related to the asymptotic 
behavior of the DGLAP evolution 
discovered many years ago \cite{Rujula}.

The study of Ref. \cite{BF1} was extended in Ref. \cite{Munich,Q2evo,HT}
to include the finite parts of anomalous dimensions
of Wilson operators and Wilson coefficients\footnote{
In the standard DAS approximation \cite{Rujula} only the singular
parts of the anomalous dimensions are used },
which predicts \cite{Q2evo,HT} the small $x$ asymptotic
form of parton distributions (PD)
in the framework of the DGLAP equation starting at some $Q^2_0$ with
the flat function:
 \begin{eqnarray}
f_a (Q^2_0) ~=~
A_a ~~~~(\mbox{hereafter } a=q,g), \label{1}
 \end{eqnarray}
where $f_a$ are the parton distributions multiplied by $x$
and $A_a$ are unknown parameters to be determined from data.

>From now on, we refer to the approach of Ref. \cite{Munich,Q2evo,HT} as
{\it generalized} DAS approximation. In generalized
DAS the flat initial conditions writen in Eq. (\ref{1}) determine,
as in the standard case \cite{BF1}, the
basic role of the anomalous dimensions singular parts while
the contribution from finite parts 
of anomalous dimensions and also from Wilson coefficients can be
considered as corrections which however
are important to have better agreement with experimental data \cite{Q2evo}.
In the present work, similarly to  Refs. \cite{BF1}-\cite{HT},
the contribution from the non-singlet quark component was neglected.

The usage of the flat initial condition given in Eq. (\ref{1}) is
supported by the actual experimental situation: low-$Q^2$ data
\cite{NMC,H197,lowQ2,Surrow} are well described for $Q^2 \leq 0.4$ GeV$^2$
by Regge theory with Pomeron intercept
$\alpha_P(0) \equiv \lambda_P +1 =1.08$,
closed to the standard ($\alpha_P(0) =1$) one. 
The small rise observed in HERA data \cite{H197,H101,lowQ2,lowQ2N,Surrow}
at low $Q^2$ can be naturally explained by including
higher twist terms (see \cite{HT,BartelsHT}).
Moreover, HERA data \cite{H197,H101,lowQ2,lowQ2N}
with $Q^2 > 1$ GeV$^2$ are in good agreement with the predictions from
the GRV parton densities \cite{GRV} 
which supports our aim to develop an
analytical form for the parton densities at small $x$ because,
at least conceptually, our method
is very closed to the GRV approach.\\

The purpose of this work is to extend the 
study of Ref. \cite{Q2evo} to compare the predictions from
the generalized DAS approach with the new precise H1 data
\cite{H1slo} for the $F_2$ $x$ slope.

The article is organized as follows.  In Section 2 we 
address the present situation with experimental data for
the slope $d\ln F_2/d\ln(1/x)$ and we shortly review 
some approaches to describe them.
Sections 3 and 4 contain, for completness, a compilation of the
basic formulae in the generalized DAS aproximation from Ref. \cite{Q2evo}
needed for the present study.
In Section 5 we compare our predictions for the derivative
$d\ln F_2/d\ln(1/x)$ with experimental data and discuss the obtained
results.

\section{The slope $d \ln F_2/d\ln (1/x)$:
experimental data and QCD phenomenology} \indent

Various groups have been able to fit the available data (mostly
separating the low and high $Q^2$ regions) using a hard input at small $x$,
$x^{-\lambda},~\lambda >0$. This is clearly different from the flat input
in the DAS approach of Refs. \cite{BF1}-\cite{HT}, also describing
reasonably well the experimental results.
In some sense it is not very surprising because
modern HERA data (at large $Q^2$)
cannot distinguish between the behavior from a steep input
parton parameterization at quite large $Q^2$, and the
steep form adquired after dynamical evolution from a flat initial
condition at quite low $Q^2$ values.
 
Moreover, for the $Q^2$ evolution based on
the full set  of anomalous dimensions 
obtained at $x \to 0$ in Ref. \cite{GSFM}
within the BFKL formalism \cite{BFKL}, the results weakly 
depend on the form of the initial condition \cite{Catani},
preserving the steep ones and changing the flat ones.
In the case of working with anomalous dimensions at fixed order in $\alpha_s$,
the initial conditions are important when the data are considered in a wide 
$Q^2$ range and it is necessary to choose adecuately the form of the PD
asymptotics at some $Q^2_0$.

As we have already discussed in the introduction, the usage of a flat
initial condition leads to the (generalized) DAS
approximation \cite{Munich}-\cite{HT}
(see the compilation of the most important
formulas in the following two sections).
An alternative to this is the choice of a steep initial condition 
at some quite large $Q^2_c$: $f_a (x,Q^2_c) \sim x^{-\lambda_c}$
(subindex {\rm c} stays for constant),
that leads to the following $Q^2$-dependence for  $f_a (x,Q^2)$
\cite{LoYn}-\cite{Regge} (if $x^{-\lambda_c} >> Const$):
 \begin{eqnarray}
\frac{f_a (x,Q^2)}{f_a (x,Q^2_c)}  
~\sim~ \frac{M_a^+(1+\lambda_c,Q^2)}{M_a^+ (1+\lambda_c,Q^2_c)},
\label{1d}
\end{eqnarray}
where $M_a^+(1+\lambda_c,Q^2)$ is the analytical continuation
(from integer $n$ to real $1+\lambda_c$ values)
of the '$+$' component of the Mellin moment of $f_a (x,Q^2)$:
 \begin{eqnarray}
M_a(n,Q^2) ~=~ \int^1_0 dx x^{n-2}
f_a (x,Q^2)
\label{1d1}
 \end{eqnarray}

When $x^{-\lambda_c} >> Const$,
the slope $\lambda_c$ should be $Q^2$-independent \cite{VoKoMa,YF93}
and the whole $Q^2$-dependence for  $f_a (x,Q^2)$
comes from the factor  $\sim M_a^+(1+\lambda_c,Q^2)$ in front of
$x^{-\lambda_c}$ in Eq. (\ref{1d}) .
Approximations similar to Eq. (\ref{1d}) have been successfully applied in
studying the $Q^2$-dependence of HERA data at large $Q^2$ values
(see Ref. \cite{Yndu} and references therein).\\

Considering separately the low $Q^2$ region, it is also possible to have
a good agreement between $F_2$ data and its Regge-like 
behavior \cite{Surrow}. 
Indeed, $F_2$ at $Q^2 \to 0$ can be extracted from the relation:
 \begin{eqnarray}
F_2 ~=~ 
\frac{Q^2}{4\pi \alpha_{em}} \sigma_{tot}^{\gamma^* p},  
\label{2d}
 \end{eqnarray}
where $\alpha_{em}$ is electromagnetic coupling constant and
$\sigma_{tot}^{\gamma^* p}$ is the total (virtual) photoproduction
cross-section.

A large amount of experimental data on hadronic cross-sections for
many different processes shows an universal rise at large energies
that gives the possibility to parameterize all these cross-sections 
as the sum of two different components
 \begin{eqnarray}
\sigma_{tot}^{\gamma^* p} ~=~ A_P s^{\alpha_P(0)-1} + A_R s^{\alpha_R(0)-1}, 
\label{3d}
 \end{eqnarray}
being $s$ the center of mass energy squared. The constants
$A_P$ and $A_R$ are process-dependent magnitudes
and the intercepts
$\alpha_P(0) \simeq 1.08$ and $\alpha_R(0) \simeq 0.5$ (see \cite{DoLa})
are universal process-independent constants. The first and second terms 
in Eq. (\ref{3d}) correspond to (soft) Pomeron and Reggeon exchange,
respectively.

>From Eqs. (\ref{2d}) and (\ref{3d}) one immediately obtains that for 
$Q^2 \to 0$ 
 \begin{eqnarray}
F_2 (x,Q^2) ~\sim~ x^{-\ep}
~~~\mbox{ and, hence, }~~~
f_a (x,Q^2)~\sim~ x^{-\ep}
~~~(\ep = \alpha_P(0)-1 \approx 0.08),
\nonumber
 \end{eqnarray}
because $s=Q^2/x$ at small $x$.\\

There were many attempts to study the whole $Q^2$ region in the framework
of Regge-asymptotics (see, for example, the reviews in Ref. \cite{CoDeRo}).  
As the reports in Ref. \cite{CoDeRo} contain quite numerous sets of models,
we will restrict ourselves to discuss below only two of them.

In Ref. \cite{Abramo}
the fit to $F_2$ experimental data 
has been done under the approach 
 \begin{eqnarray}
f_a (x,Q^2) \sim x^{-\lambda (Q^2)}
\label{3da}
 \end{eqnarray}
and a fast changing of the $\lambda (Q^2)$
has been found in the transition range $Q^2 \sim 5\div$10 GeV$^2$.
Unfortunately, it is rather difficult to reconcile
for the whole $Q^2$ range
the Regge-like behavior given by Eq. (\ref{3da}) with DGLAP evolution.
Some progress along this line has been done in 
Ref. \cite{Regge} that was also based on
the flat initial conditions as given by Eq. (\ref{1}).
However, in Ref. \cite{Regge} the PD structure is limited by
the Regge-like form of Eq. (\ref{3da}),
permiting to recoincile it with DGLAP evolution
only separately at low $Q^2$, where $\lambda (Q^2)$ is close to $0$
(or to $\ep$), and at large $Q^2$, where $\lambda (Q^2) \sim \lambda_c$.
The structure function $F_2$  and parton distributions
have been obtained in Ref. \cite{Regge} for the whole $Q^2$ range
only as a combination of these two representations.

For other types of
models (see \cite{CaKaMeTTV,DeJePa}) the phenomenological 
$Q^2$-dependence of $\lambda (Q^2)$ has been introduced in the form:
 \begin{eqnarray}
\lambda (Q^2) ~=~ \ep \left(1 + \frac{Q^2}{Q^2 + c} \right)
\nonumber
 \end{eqnarray}
with a fitted constant $c$, that produces the soft values of the
slope $\lambda (Q^2)$ close to $\ep$ 
at low $Q^2$ and the hard ones
$\lambda (Q^2) \sim \lambda_c \sim 0.2\div 0.3$ at
$Q^2 \geq 20$ GeV$^2$.\\

Very recently new precise experimental data on $\lambda (Q^2)$ 
has become available \cite{H1slo}.
The H1 data points are shown in Fig. 1 where one can observe that 
for fixed $Q^2$, $\lambda$ is independent on $x$ 
within the experimental uncertainties, in the range $x <0.01$.
Indeed, H1 data are well
described by the power behavior \cite{H1slo}:
\begin{eqnarray}
F_2 (x,Q^2) ~=~ C x^{-\lambda (Q^2)},
\label{1dd}
\end{eqnarray}
where $\lambda (Q^2) = \hat a \ln(Q^2/\Lambda^2)$ with $C \approx 0.18,
\hat a \approx 0.048$ and $ \Lambda =292$ MeV.
The linear rise of $\lambda$ with $\ln Q^2$ given by Eq. (\ref{1dd})
is plotted in Fig 2.

A similar analysis for extracting $\lambda (Q^2)$ as a
function of $x$ has been carried out by the ZEUS Collaboration. As it
is possible to see in Fig. 8 of Ref. \cite{Surrow}, the ZEUS data for 
$\lambda (Q^2)$ are compatible with a constant value $\sim 0.1$ at $Q^2 <
0.6$ GeV$^2$, as it is expected under the assumption of single soft Pomeron
exchange within the framework of Regge phenomenology.
For the case of H1, this behavior can also be inferred
from the new preliminary H1 data \cite{DIS02} at quite low
values of $Q^2$.
%
%
\\

Let us point out that, even though our results in the framework
of the generalized DAS approximation  (Eqs. (\ref{9.10})-(\ref{9}) below)
do not have an explicit power-like behavior, they really mimic a power
law shape over a limited region of $x$ and $Q^2$ (see Section 4).
In adition we have observed earlier \cite{Q2evo} that
the $x$ dependence of the effective slopes in generalized DAS
is not strong and the $F_2$ effective slope was in good agreement
with old (less precise) H1 data \cite{H197}.
We repeat in Section 5 the analysis performed in Ref. \cite{Q2evo}
but now using the new precise H1 data for the slope \cite{H1slo}.

\section{$Q^2$ dependence of $F_2$ and parton distributions
in generalized DAS approximation} \indent

To begin with, we shortly write down  the results of the generalized 
DAS firstly presented in Ref. \cite{Q2evo}.
The small $x$ behavior of the parton densities and
$F_2$ at NLO has the form\footnote{
Hereafter $z=x/x_0$, where
$x_0$ is a free parameter which limits the applicability range of 
formulae (\ref{9.10})-(\ref{9})
and can be fitted from experimental data together with the magnitudes
of gluon and sea quark distributions at $Q^2_0$. As it has been
shown in Ref. \cite{Q2evo}, the fits to $F_2$ HERA data depend very
slightly on the concrete $x_0$ value.}
:
\begin{eqnarray}
f_a(z,Q^2) &=& f_a^+(z,Q^2) + f_a^-(z,Q^2) ~~~~~\mbox{ and }~ \nonumber \\
f_a^-(z,Q^2) &\sim &
\exp(- d_{-}(1)s-D_{-}(1)p) ~+~O(z) 
\label{9.10} \\
f_g^+(z,Q^2) &\sim &
I_0(\sigma) \exp(- \overline d_{+}(1)s-\overline D_{+}(1)p)
~+~O(\rho) 
\label{9.11} \\
f_q^+(z,Q^2) &\sim &
f_g^+(z,Q^2) \cdot
\biggl[ (1 - \bar{d}_{+-}^q(1) a_s(Q^2)) 
\frac{\rho I_1(\sigma)}{I_0(\sigma)}
      + 20 a_s(Q^2)  \biggr]
%
+O(\rho) \label{9.12} \\
F_2(z,Q^2)&=& e \cdot \biggl(f_q(z,Q^2) + \frac{2}{3}f a_s(Q^2) f_g(z,Q^2)
 \biggr),
\label{9}
\end{eqnarray}
where $e=\sum_i^f e_i^2/f$ is the average charge square of $f$ effective quarks,
$a_s=\alpha_s/(4\pi)$ and
 \begin{eqnarray}
s&=&\ln\left(\frac{a_s(Q^2_0)}{a_s(Q^2)}\right)~~,~~~~ 
p=a_s(Q^2_0)-a_s(Q^2)~~,~~~~
D_{\pm}= d_{\pm\pm}-\frac{\beta_1}{\beta_0}d_{\pm}~~,\nonumber \\
\sigma &=& 2\sqrt{(\hat d_{+}s+\hat D_{+}p)\ln z} ~~, ~~~~
\rho = \sqrt{\frac{(\hat d_{+}s+\hat D_{+}p)}{\ln z}}=
\frac{\sigma }{2\ln(1/z)}
~~,
\label{a1}
\end{eqnarray}
and $\beta_0$ and $\beta_1$ are first two terms of QCD $\beta$-function.

%
The components of the leading order (LO) AD $d_{-}(n)$ and 
singular parts $\hat d_{+}$ and regular ones 
$\overline d_+(n)$ of the LO AD
$d_+(n)=\hat d_{+}/(n-1) + \overline d_{+}(n)$ 
have the following form (at $n \to 1$):
 \begin{eqnarray}
\hat d_{+} &=& -\frac{12}{\beta_0} ~~, ~~~
\overline d_{+}(1) = 1+\frac{20 f}{27\beta_0}~~,~~~
d_{-}(1) = \frac{16 f}{27\beta_0} \label{9.03}
 \end{eqnarray}

The corresponding NLO components 
can be represented as follows:
 \begin{eqnarray}
\hat d_{++} &=& \frac{412}{27\beta_0}f ~~, ~~~
\hat d^q_{+-} = -20 ~~, ~~~
\hat d^g_{+-} = 0~~, \nonumber \\
\overline d_{++}(1) &=& \frac{8}{\beta_0}
\biggl( 36 \zeta_3 + 33 \zeta_2 - \frac{1643}{12} +\frac{2}{9}f 
\Bigr[ \frac{68}{9} -4 \zeta_2 - \frac{13}{243}f \Big] \biggr)~~, \nonumber \\
\overline d^q_{+-}(1) &=& 
\frac{134}{3} -12 \zeta_2 - \frac{13}{81}f 
~~, ~~~
\overline d^g_{+-}(1) = \frac{80}{81}f~~, \nonumber \\
d_{--}(1) &=& \frac{16}{9\beta_0}
\biggl( 2 \zeta_3 - 3 \zeta_2 + \frac{13}{4} + f 
\Bigr[  4 \zeta_2 - \frac{23}{18} + \frac{13}{243}f \Big] \biggr)~~,
\nonumber \\
d^q_{-+}(1) &=& 0~~, ~~~
d^g_{-+}(1) = -3 \Bigl( 1+ \frac{f}{81} \Bigr)~~.
 \label{9.3}
 \end{eqnarray}

Some interesting features of the results given in Eqs. (\ref{9.10})-({\ref{a1})
are summarized below:
 
\begin{itemize}
\item Both, the gluon and quark singlet densities given above are presented in
terms of two components ('$+$' and '$-$') which are obtained from the
analytical $Q^2$-dependent expressions of the corresponding ('$+$' and 
'$-$') components of PD
moments.

\item The '$-$' component is constant at small $x$, whereas the '$+$'
component grows at $Q^2 \geq Q^2_0$ as $\sim \exp (\sigma)$, where
$\sigma $ contains the positive LO term  
$|\hat d_{+}|s\ln(1/z)$ and the negative NLO one
$|\hat D_{+}|p\ln(1/z)$ (see Eq. (\ref{a1})). So, the most important part from the
NLO corrections (i.e. the singular part at $x \to 0$) is taken in a proper way: it comes 
directly into the argument of the Bessel functions and does not spoil
the applicability of perturbation theory at low $x$ values.
\end{itemize}

\section{$Q^2$ dependence of the slope $d \ln F_2/d\ln (1/x)$
in generalized DAS
approximation} \indent

The behavior of the parton densities and the structure
funcion $F_2$ within generalized DAS approach, given by
Eqs. (\ref{9.10})-(\ref{9}), can be
represented by a power law shape over a limited region of $x$ and
$Q^2$ \cite{Q2evo,HT}:
 \begin{eqnarray}
f_a(x,Q^2) \sim x^{-\lambda^{eff}_a(x,Q^2)}
 ~\mbox{ and }~
F_2(x,Q^2) \sim x^{-\lambda^{eff}_{F2}(x,Q^2)}
\nonumber    \end{eqnarray}

Because $d/d\ln x = d/d\ln z$,
the effective slopes can be obtained directly from Eqs. 
(\ref{9.10})-(\ref{9}). They
have the form:
 \begin{eqnarray}
\lambda^{eff}_g(z,Q^2) &=& \frac{f^+_g(z,Q^2)}{f_g(z,Q^2)} \cdot
\rho \cdot \frac{I_1(\sigma)}{I_0(\sigma)}
\nonumber
\\
\lambda^{eff}_q(z,Q^2) &=& \frac{f^+_q(z,Q^2)}{f_q(z,Q^2)} \cdot
\rho \cdot \frac{ I_2(\sigma) (1- \overline d^q_{+-}(1) a_s(Q^2))
 + 20 a_s(Q^2) I_1(\sigma)/\rho}{ I_1(\sigma) 
(1- \overline d^q_{+-}(1) a_s(Q^2))
 + 20 a_s(Q^2) I_0(\sigma)/\rho}
\label{10.1}
\\
\lambda^{eff}_{F2}(z,Q^2) &=& \frac{\lambda^{eff}_q(z,Q^2) \cdot
f^+_q(z,Q^2) + (2f)/3a_s(Q^2)\cdot \lambda^{eff}_g(z,Q^2) \cdot
f^+_g(z,Q^2)}{f_q(z,Q^2) + (2f)/3 a_s(Q^2)\cdot f_g(z,Q^2)}
\nonumber
\end{eqnarray}

It is interesting to emphasize that from Eq. (\ref{10.1}) one obtains
that the gluon effective slope $\lambda^{eff}_g$ 
is larger than the quark slope $\lambda^{eff}_q$ \cite{Q2evo},
which is in excellent agreement with 
MRS \cite{MRS} and GRV \cite{GRV} analyses (see also Ref. \cite{fits}). 

On the other hand the effective slopes $\lambda^{eff}_a $ and 
$\lambda^{eff}_{F2}$ in Eq. (\ref{10.1}) depend on the magnitudes
of the initial PD, $A_a$, and also on the chosen input values $Q^2_0$
and $\Lambda$. However, at quite large $Q^2$, where the ``$-$'' component
is negligible, the dependence on the initial PD disappears, having
in this case, for the asymptotic behavior, the following
expressions\footnote{
The asymptotic formulae given in Eq. (\ref{11.1})
work quite well at any $Q^2 \geq Q^2_0$ values,
because at $Q^2=Q^2_0$ the values of
$\lambda^{eff}_a $ and $\lambda^{eff}_{F2}$ are equal zero. 
The use of approximations in Eq. (\ref{11.1}) instead of the exact results 
given in Eq. (\ref{10.1}) underestimates 
(overestimates) only slightly the gluon (quark) slope
at $Q^2 \geq Q^2_0$.
For the $F_2$ case, the similarity of $\lambda^{eff}_{F2} $ and 
$\lambda^{eff,as}_{F2}$ values is shown in Fig 1.}:
 \begin{eqnarray}
\lambda^{eff,as}_g(z,Q^2) &=& 
\rho\, \frac{I_1(\sigma)}{I_0(\sigma)} \approx \rho - 
\frac{1}{4\ln{(1/z)}} 
\nonumber \\
\lambda^{eff,as}_q(z,Q^2) &=& 
\rho \cdot \frac{ I_2(\sigma) (1- \overline d^q_{+-}(1) a_s(Q^2))
 + 20 a_s(Q^2) I_1(\sigma)/\rho}{ I_1(\sigma) 
(1- \overline d^q_{+-}(1) a_s(Q^2))
 + 20 a_s(Q^2) I_0(\sigma)/\rho}
 \nonumber \\
&\approx & \biggl( \rho - \frac{3}{4\ln{(1/z)}} \biggr) \biggl(1- 
\frac{10a_s(Q^2)}{(\hat d_+ s + \hat D_+ p)} \biggr)
\label{11.1} \\
\lambda^{eff,as}_{F2}(z,Q^2) 
&=& \lambda^{eff,as}_q(z,Q^2) 
\frac{ 1 + 6 a_s(Q^2)/\lambda^{eff,as}_q(z,Q^2)}{ 1 + 
6 a_s(Q^2)/\lambda^{eff,as}_g(z,Q^2)} + ~O(a_s^2(Q^2)) 
\nonumber \\
&\approx & 
\lambda^{eff,as}_q(z,Q^2) + \frac{3 a_s(Q^2)}{\ln(1/z)},
\nonumber
\end{eqnarray}
where the symbol $\approx $ denotes that an approximation was done in the 
expansion of the modified Bessel functions $I_n(\sigma)$ $(n=0,1,2)$. 
These approximations are
accurate only at large $\sigma $ values (i.e. at large $Q^2$
and/or small $x$).

Finally, we note that at LO the $F_2$ slope
$\lambda^{eff,as}_{F2}$ is equal to the quark slope
$\lambda^{eff,as}_q$ and it coincides, for very large values
of $\sigma $, with the result from Ref. \cite{Navelet} in the case of a
flat input (see also the first article in Ref. \cite{CoDeRo}). 
At NLO, $\lambda^{eff,as}_{F2}$ lies between the quark and gluon 
slopes but closer to the former (see Fig. 3 in Ref. \cite{Q2evo}).

\section{Comparison with experimental data
} \indent

Using the results of previous section we have
analyzed  HERA data for the slope $d\ln F_2/d\ln (1/x)$
at small $x$ from the H1 Collaboration\footnote{
In this work we only use H1 data \cite{H1slo}.
The preliminary ZEUS data for the slope $d\ln F_2/d\ln (1/x)$
are only available through points in Figs. 8 and 9 of Ref. \cite{Surrow}.
They shown quite similar properties in comparison with H1 data 
\cite{H1slo}. Unfortunately, the ZEUS numerical values are
yet unavailable and we can not analyze them in the present work.}
\cite{H1slo}.

Initially our results for $\lambda^{eff}_{F2}$ 
depends on five parameters, i.e. $Q_0^2$, $x_0$, $A_q$, $A_g$ and 
$\Lambda_{\ms}$(f=4).
In our previous article \cite{Q2evo}
we have fixed $\Lambda_{\ms}$(f=4)=250 MeV which
was a reasonable value extracted from the traditional (higher $x$)
experiments.
All other parameters have been fitted and good agreement
with $F_2$ HERA data has been
found for $Q_0^2\sim 1$ GeV$^2$ (all results depend very slightly
on $x_0$).

 In this work we take $\Lambda_{\ms}$(f=4)=292 MeV in agreement
with the more recent H1 results \cite{H1slo}
and other analyses (see Ref. \cite{KriKo} and references therein)
and we fit directly the slope $d\ln F_2/d\ln (1/x)$ data using
Eq.(\ref{10.1}). The result is shown in Fig. 1.
For comparison we also plot the curves
from a fit to $F_2$ data in Ref. \cite{Q2evo}
where the value 250 MeV was used.
The results are very similar and demonstrate the very important
feature of an approximate $x$-independence of
$\lambda^{eff}_{F2}$ as given by Eq. (\ref{10.1}), which is in
agreement with H1 data \cite{H1slo}. 

We also present in Fig. 1 the asymptotic values for the slope
$\lambda^{eff,as}_{F2}$ as obtained from Eq. (\ref{11.1}).
The agreement with data and with the other curves
is also rather good if one takes in consideration that,
in this case, there is no fit involved because the only parameters
entering Eq. (\ref{11.1}) are the fixed values
$\Lambda_{\ms}$(f=4)=292 MeV and $x_0=1$.

Thus, the tiny $x$ dependence of the slope $\lambda^{eff}_{F2}(x,Q^2)$
in the considered region of $x$ and $Q^2$ support the
possibility to successfully use our generalized DAS approximation
in the type of $x$ independent analysis for the $F_2$ slope.

Fig. 2 shows the experimental data for $\lambda^{eff}_{F2}$
and the corresponding H1 parameterization \cite{H1slo}
written above in Eq. (\ref{1dd}).
We have also plotted the 
result from Eq. (\ref{11.1}) and one 
from Eq. (\ref{10.1}) using the parameters from our previous article
\cite{Q2evo} as it was presented before in Fig. 1.
For both cases, we give it for two representative $x$ values.

>From the visual inspection of Fig. 1, one can notice that
the boundaries and mean values of the experimental $x$ ranges 
\cite{H1slo} increase proportionally with $Q^2$, what is related
to the kinematical restrictions in the HERA experiments:
$x \sim 10^{-4} \cdot Q^2$
(see Refs. \cite{H197,H101,ZEUS01,lowQ2} and, for example,
Fig. 1 of \cite{Surrow}).

Fig. 3 shows H1 experimental data \cite{H1slo}
for $\lambda^{eff}_{F2}$
and the H1 parameterization (Eq. (\ref{1dd})), as given in Fig. 2, but
this time in comparison with the asymptotic
values $\lambda^{eff,as}_{F2}$ calculated from Eq. (\ref{11.1})
using $x = a \cdot 10^{-4} \cdot Q^2$
with  $a=0.1,1$ and $10$.
One has a reasonable agreement with H1 data for $Q^2 > 2$ GeV$^2$
using $a$ between $0.1$ and $1$ (the two lower dashed curves in Fig. 3),
which approximately
corresponds to the middle points of the measured $x$ range.

\section{Conclusions} \indent

We have studied the $Q^2$ dependence of the slope 
$\lambda^{eff}_{F2}=d\ln F_2/d\ln (1/x)$ at small $x$ in the 
framework of perturbative QCD. Our results are in good agreement with 
new precise experimental H1 data \cite{H1slo} at $Q^2 \geq 2$ GeV$^2$,
where perturbative theory can be applicable.

Despite the fact that our approach, which can be called generalized
doubled asymptotic scaling approximation, is based on pure perturbative 
grounds: a flat initial conditions at $Q^2_0 \approx 1$ GeV$^2$
and dynamical evolution to $Q^2 \geq Q^2_0$ (it is very close conceptually
to GRV approach but contains exact analytical $Q^2$ evolution), it can be
reasonable applied for the new precise data of the 
slope $\lambda^{eff}_{F2}$.

The agreement between $\lambda^{eff}_{F2}$ data
and perturbative QCD has been already observed by H1\cite{H101} and ZEUS
\cite{Surrow} collaborations. The obtained linear
rise of $\lambda (Q^2)$ with $\ln Q^2$ (see for example Figs. 2 and 3),
parametrized by H1 as in Eq. (7),
can be naively interpreted in the strong nonperturbative way, i.e.
$\lambda (Q^2) \sim 1/\alpha_s(Q^2)$. Our analysis, however, demonstrates
that the rise can be explained as $\sim \ln\ln Q^2$, that is natural in 
perturbative QCD at low $x$ (see \cite{Rujula}, \cite{BF1}-\cite{HT}
and references therein):
when the coupling constant is running, the renormalization group
leads to the small $x$ behavior for the PD $\sim \ln(\alpha_s(Q^2))$ at
the LO of perturbation theory and $\sim \alpha_s(Q^2)$ at NLO
(see Eqs. (\ref{9.10})-(\ref{a1}) and discussions after Eq. (\ref{9.3})).

The good agreement between perturbative QCD and the experiment obtained
here and in Ref. \cite{Q2evo,HT} demonstrate the fact that for
$Q^2 > 2$ GeV$^2$ non-perturbative contributions as shadowing
effects \cite{Levin},  higher twist effects
\cite{Bartels} and others seem to be quite small 
(see also Ref. \cite{BaGoPe} and references therein) 
or they seem to be canceled
between them and/or with $\ln(1/x)$ terms contained in the higher orders
of perturbation theory. 
Note, however, that higher twist corrections are important at 
$Q^2 \leq 1$ GeV$^2$, as it has been demonstrated in
Ref. \cite{HT,BartelsHT,KriKo}.
To clear up the correct contributions from nonperturbative
dynamics and higher orders containing strong $\ln(1/x)$ terms,
it is necessary
further efforts in the development of theoretical
approaches.

Moreover, the good agreement between perturbative QCD and experimental
data at low $Q^2$ 
can be explained with a larger effective scale for
the QCD coupling constant \cite{Q2evo,HT}.
The similar behavior has already been observed in the framework of
perturbative QCD \cite{DoShi} and in BFKL-motivated approaches 
\cite{bfklp}-\cite{Salam} (see the recent review in Ref. \cite{Andersson}
and discussions therein).

Note that large NLO corrections
calculated recently in the framework of BFKL \cite{FaLi}
(see also \cite{KoLi}) lead to a strong suppression of the LO
BFKL results for the high energy
asymptotic behavior of the cross-section
(see, for example, \cite{bfklp} an \cite{bfklp1}). 
The careful inclusion of NLO corrections leads to results which are
quite close to those obtained in pure perturbative QCD \cite{bfklp1}.
This fact can give an additional support for the good applicability
of perturbation theory in the small $x$ range, where, as was expected
before, nonperturbative effects should give an essential contribution.

As the next step, it could be very useful to apply
the generalized doubled asymptotic scaling approach to make a combined
analysis of HERA data for $F_2$, $dF_2/d\ln(Q^2)$, $d\ln F_2/d\ln(1/x)$ 
and  $F_L$.
We hope to consider this work in a forthcoming article,
including higher twist corrections in
the $Q^2$-evolution approach given by Eqs. (\ref{9.10})-(\ref{9}).
We also plan to consider additional terms in the initial condition
given by Eq. (\ref{1}) 
proportional to $\ln(1/x)$ and $\ln^2(1/x)$.

We hope that the analysis will be relevant to find
out the kinematical region where
the well-established perturbative QCD formalism
can be safely applied at small $x$. Moreover, the study should clear up
the reason of the good agreement between the small $x$ relation of $F_L$, 
$F_2$ and $dF_2/d\ln(Q^2)$ obtained in pure perturbative QCD in
Ref. \cite{KoPaFL} (and based on previous works \cite{method,KoPaG}), 
the experimental data for these structure
functions \cite{H1FL,H101} and the predictions of Ref. \cite{KoLiZo} 
in the framework of $k_t$-factorization \cite{GSFM,Catani,CCH}.

\vspace{1cm}
\hspace{1cm} \Large{} {\bf Acknowledgements}    \vspace{0.5cm}

\normalsize{}

The study is supported in part by the RFBR grant 02-02-17513
and  by the Heisenberg-Landau program. 
One of the authors (A.V.K.) was supported in part
by Alexander von Humboldt fellowship and INTAS  grant N366.
G.P. acknowledges the support of Galician research funds
(PGIDT00 PX20615PR) and Spanish CICYT (FPA2002-01161).

\newpage

\vspace{0.5cm}

\hspace{1cm} {\Large{\bf Figure captions}}    \vspace{0.5cm}

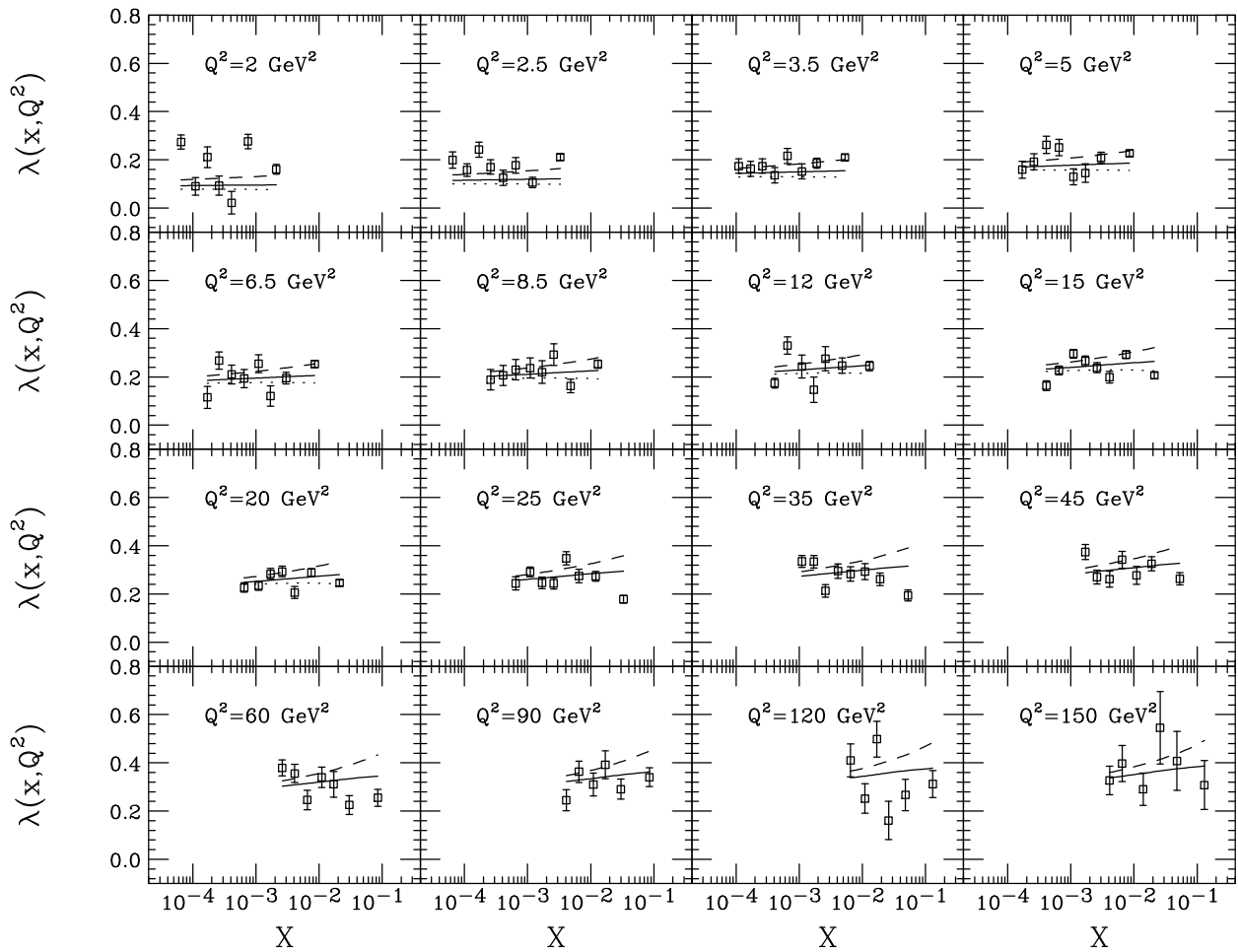
{ \bf Figure 1.} 
The derivative $d \ln F_2/d\ln (1/x)$ (effective slope $\lambda$ )
as a function of $x$ for different $Q^2$ values. 
Data points are from H1 \cite{H1slo}. Only statistical uncertainties are
shown
The solid line is the result from a fit using $\lambda^{eff}_{F2}$
in Eq. (\ref{10.1}) with fixed $Q_0^2=1$ GeV$^2$ and $x_0=1$. 
The dotted line is the same as the solid but with the parameters
from a fit to $F_2$ data in Ref. \cite{Q2evo}. 
The dashed line corresponds to the asymptotic expression
$\lambda^{eff,as}_{F2}$ in Eq. (\ref{11.1}).

\vspace{0.5cm}

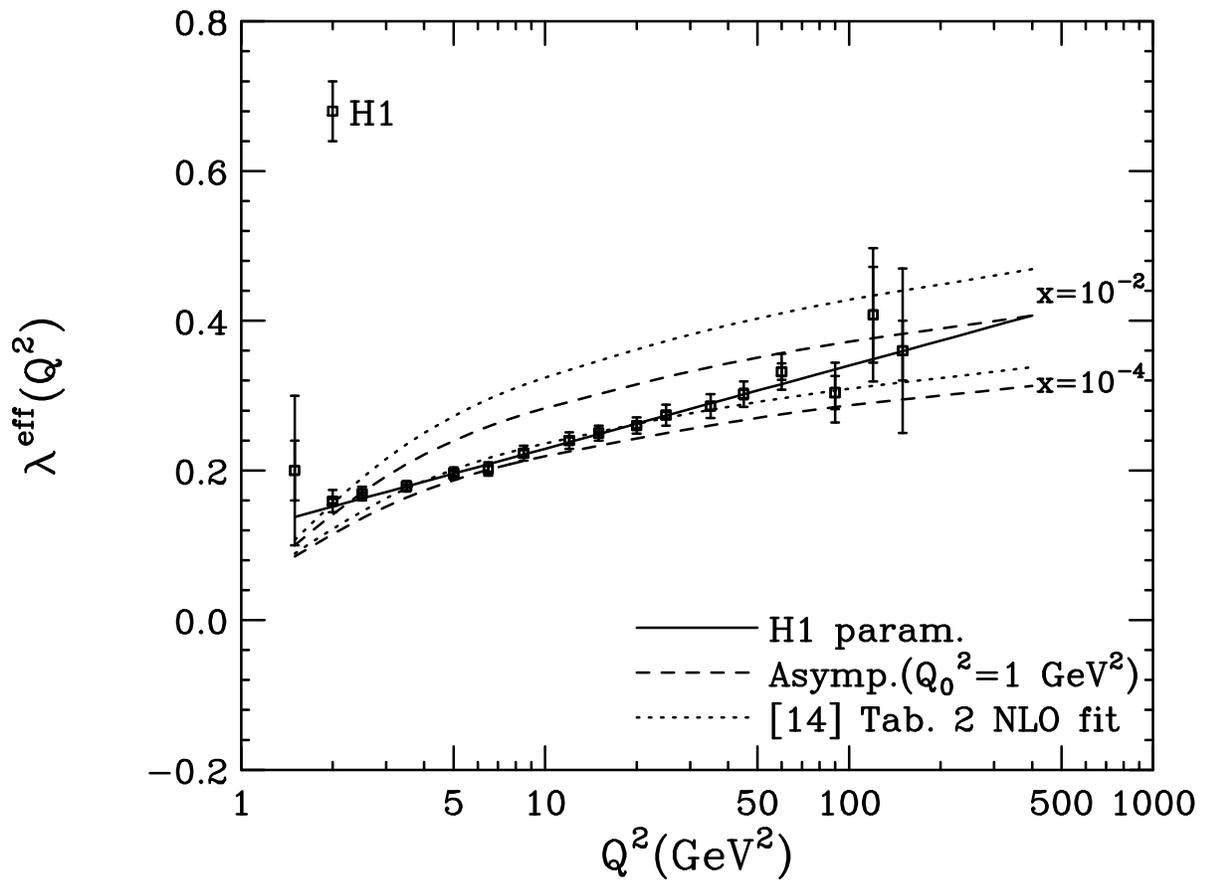
{ \bf Figure 2.} 
The derivative $d \ln F_2/d\ln (1/x)$ (effective slope $\lambda$ )
as a function of $Q^2$. 
Data points are from H1 \cite{H1slo}. 
Outer error bars include statistic and systematic added in quadrature.
Inner bars correspond to statistic errors.
The solid line corresponds to the H1 parameterization \cite{H1slo}
given in Eq. (7).
Dotted and dashed curves are produced as in Fig. 1.
For the lower (upper) curves, the value $x = 10^{-4}$ ($x = 10^{-2}$) was
used.

\vspace{0.5cm}

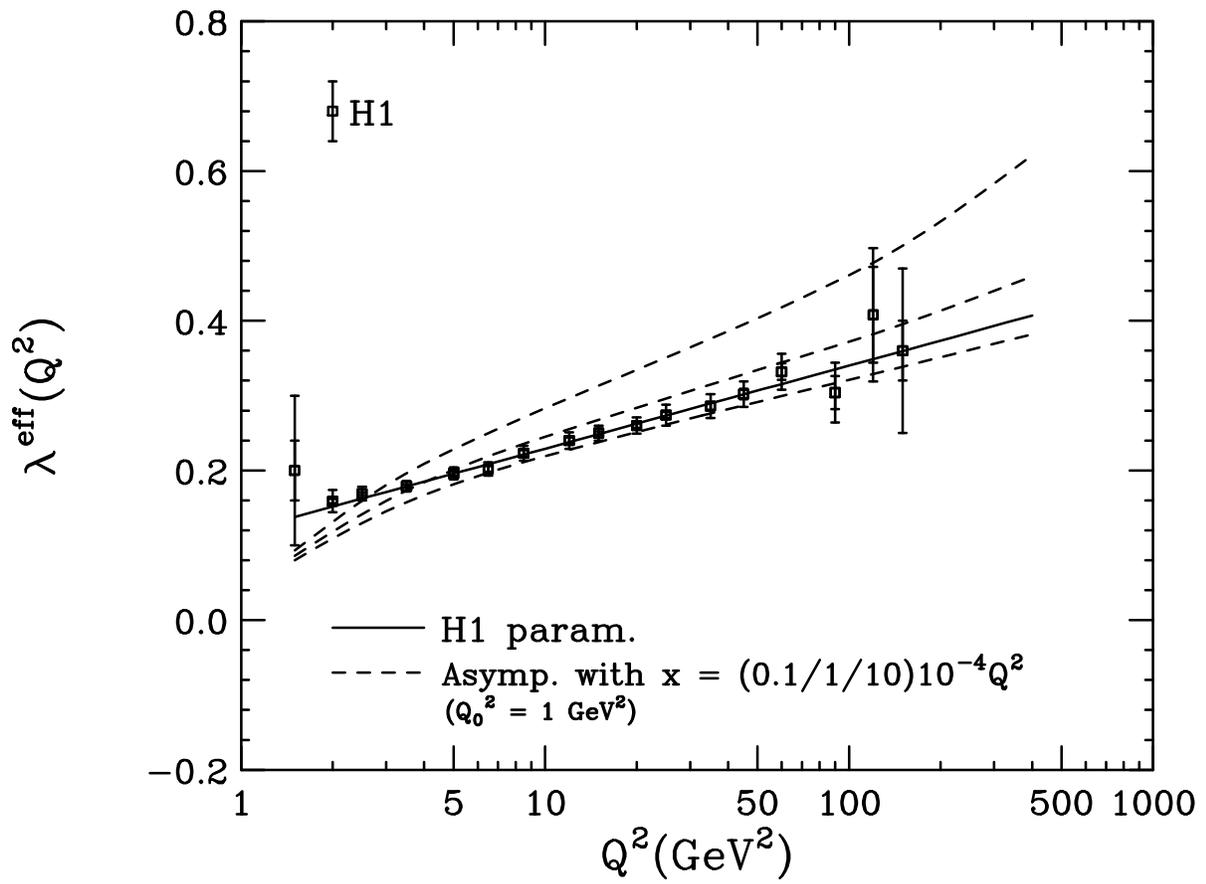
{ \bf Figure 3.} 
The derivative $d \ln F_2/d\ln (1/x)$ (effective slope $\lambda$ )
as a function of $Q^2$. 
Data points are from H1 \cite{H1slo}. Error bars and solid line are as
indicated in Fig. 2.
The dashed lines were calculated with Eq. (\ref{11.1})
using $x = a \cdot 10^{-4} \cdot Q^2$
with $a=0.1,1$ and $10$. Upper curves correspond to larger $x$.

\end{document}